%Paper: hep-ph/9508239
%From: Jan Czyzewski <czyzewsk@hef.kun.nl>
%Date: Fri, 4 Aug 95 19:11:05 METDST

\def\PRep{Phys.\ Rep.\ }
\def\PR{Phys.\ Rev.\ }
\def\PRL{Phys.\ Rev.\ Lett.\ }
\def\ZP{Z.\ Phys.\ }
\def\NP{Nucl.\ Phys.\ }
\def\PL{Phys.\ Lett.\ }
\def\T{\perp}
\def\pT{p_\T}
\def\qT{q_\T}
\def\hT{h_\T}
\def\kT{k_\T}
\def\Pol{{\cal P}}
\def\vpT{\vec p_\T}
\def\vqT{\vec q_\T}
\def\vhT{\vec h_\T}
\def\vkT{\vec k_\T}
\def\vPol{\vec{\cal P}}
\magnification\magstep 1
\font\bbf=cmbx10 scaled\magstep2
\pageno=0
\footline{\ifnum\pageno=0 \hfil \else \hss\tenrm\folio\hss\fi}

\null
\vskip 3.5cm

\centerline{\bbf Single spin asymmetry in inclusive pion production,}
\vskip 0.2cm
\centerline{\bbf Collins effect and the string model}
\vskip 0.3cm
\centerline{\bbf (revised version)}
\bigskip
\bigskip

\centerline{\bf X. Artru}
\centerline{\it Institut de Physique Nucl\'eaire de Lyon,
IN2P3-CNRS et Universit\'e Claude Bernard,}
\centerline{\it F-69622 Villeurbanne Cedex, France}
\medskip

\centerline{{\bf J. Czy\.zewski}\footnote*{ On leave from \it Institute of
Physics, Jagellonian University, ul. Reymonta 4, PL-30-059 Krak\'ow, Poland }}
\centerline{\it Institute of High-Energy Physics, University of Nijmegen,}
\centerline{\it Toer\-nooi\-veld~1, NL-6525~ED Nijmegen, The Netherlands }
\medskip

\centerline{\bf H. Yabuki}
\centerline{\it Department of Mathematics, Hyogo University of Teacher
Education,}
\centerline{\it Yashiro, Hyogo, 673-14 Japan}

\vskip 1.5cm
\centerline{\bf Abstract}
\smallskip

We calculated the single spin asymmetry in the inclusive pion production in
the fragmentation region of transversely polarized proton-proton collisions.
We generated the asymmetry at the level of fragmentation function (Collins
effect) by the Lund coloured string mechanism.
We compared our results to the presently available experimental data.
We obtained a qualitative agreement with the data after assuming that the
transverse polarizations of the $u$ and the $d$ quarks in the proton are +1
and --1, respectively, at $x_B = 1$.

\vskip 1cm
{\bf TPJU 13/95}

{\bf hep-ph/9508239}

{\bf August 1995}

\eject
\noindent
%*************************************************************************
{\bf 1. Introduction}
\smallskip

\noindent
Quantum Chromodynamics predicts that single transverse spin asymmetries are
suppressed in hard collisions, as a consequence of helicity conservation
(chiral
invariance) in the subprocess.  These asymmetries indeed appear as
interferences
between helicity amplitudes which differ by one unit of helicity, therefore
they
vanish in the limit $m_{\rm quark} \to 0$, or equivalently $Q^2 \to \infty$
($Q$
measures the hardness of the subprocess).  Nevertheless, a number of high $p_T$
reactions persist in showing large asymmetries [1].

These facts do not invalidate QCD but mean that the approach to the asymptotic
regime in $\pT$ is very slow, as regards polarization.  However, in spite of
their ``nonasymptotic'' character, it is not unreasonable to think that the
mechanisms of the asymmetries lie at the parton level.  In other words, the
asymmetries would be manifestations of quark transverse spin (or
transversity).  Thus, we could extract information from them about the quark
transversity distribution in the nucleon and/or the transversely polarized
quark
fragmentation.  In this paper, we shall present a model for the spin asymmetry
in the reaction
$$
p\!\uparrow + \, p \rightarrow \pi + X
\eqno(1.1)$$
which, unlike previous approaches [2,3], involves the transverse spin asymmetry
of the polarized quark fragmentation [4,5], which hereafter will be referred to
as the Collins effect.

The paper is organized as follows:  section 2 gives a very short review of the
experimental data.  In section 3, we explain how the Collins
asymmetry can give rise to the observed single spin asymmetry in reaction (1.1)
and deduce lower bounds on the transverse polarizations of the quarks in the
proton, as well as on the size of the Collins effect.  Section 4 presents a
quantitative model based on string fragmentation and section 5 gives the
numerical results.  Section 6 contains discussion of our results and
conclusions.

\bigskip
\medskip
\noindent
%****************************************************************************
{\bf 2. Main features of single spin asymmetry in inclusive pion production}
\smallskip

\noindent
A strong polarization effect has been observed in recent years by the Fermilab
E704 collaboration in the reaction (1.1) with 200 GeV transversely polarized
projectile protons [6-8].  The asymmetry is defined as
$$A_N(x_{F}, \pT) \equiv {\sigma\uparrow - \, \sigma\downarrow
\over \sigma\uparrow + \, \sigma\downarrow},
\eqno(2.1)$$
assuming that $\uparrow$ refers to the $+ \hat y$ direction (vertical upwards),
and the transverse momentum $\vpT$ of the pion points towards the $+ \hat x$
direction ($\vec p_{\rm beam}$ is along the $\hat z$ axis).  In other words,
positive $A_N$ means that for  upward polarization, the pions tend to go to
the left.  $x_F = 2p_z^{\rm CM}/\sqrt{s}$ is the longitudinal momentum fraction
of the produced pion and $\pT$ is its transverse momentum.

\noindent
The data covered two kinematical regions:

\item{-}
Projectile fragmentation region, $x_F \ge 0.2$.  In this region the
200GeV E704 data [6-8] show large asymmetries for all pions; positive for
$\pi^+$ and $\pi^0$ and
negative for $\pi^-$.  The asymmetries vary from about 0 at $x_F\sim 0.2$ to
about +0.4, +0.15 and $-0.4$, for $\pi^+$, $\pi^0$ and $\pi^-$ respectively, at
$x_F \sim 0.7-0.8$ and $\pT > 0.7$GeV. The earlier 13.3 and 18.5GeV data [9]
showed the asymmetry
of $\pi^+$ reaching 0.1 at $x_F = 0.6$ but that of $\pi^-$ consistent with
zero.
However, that measurement was done for $\pi^+$ and $\pi^-$ in different $\pT$
regimes.

\item{-}
Central region, $x_F \sim 0$.
Ref.~[10] reported large positive asymmetry of $\pi^0$ for the transverse
momentum fraction $x_T = 2p_T/\sqrt{s} > 0.4$. However, the reanalysis of
that data [11] showed the asymmetry consistent with zero in the whole $x_T$
range covered. Former experiments, at 13.3, 18.5, 24 and 40GeV, [12] observed
significant asymmetries in the central region and high $\pT$.

\noindent
In this paper we shall concentrate only on the forward fragmentation
region.

\bigskip
\medskip
\noindent
%**************************************************************************
{\bf 3. Possible explanations of the asymmetry}

\bigskip
\noindent
{\sl 3.1 Generalities from the parton model.}

\smallskip
\noindent
In the factorized parton model, the cross section for $p+p \to \pi+X$ in the
forward hemisphere is a convolution of the parton distribution $q(x,\vqT)$, the
parton-hadron scattering cross section $\hat\sigma_{q+B\to q'+X} \equiv
\hat\sigma_{q\to q'}$ and the parton fragmentation function
$D_{\pi/q'}(z,\vhT)$.  In short-hand notations,
$$\sigma_{A \rightarrow \pi} \approx q
\otimes \hat \sigma_{q \to q'}
\otimes D_{\pi/q'}
\eqno(3.1)$$
Each
factor in this equation may or may not depend on spin.  Transverse polarization
can act at three different levels:

\item{a)} in a dependence of $q(x,\vqT)$ on the azimuth of
$\vqT$ (hereafter referred to as {\it Sivers effect} [2,13]).

\item{b)} in a single spin asymmetry in $ \hat \sigma_{q \to q'}$
hereafter referred to as {\it Szwed mechanism}
[3]. In this case, (but not necessarily in case $a$),
the quark $q$ has to inherit a part of the polarization of the proton.

\item{c)} in a dependence of $D_{\pi/q'}(z, \vhT)$ on the azimuth of $\vec
h_{\T}$ hereafter referred to as {\it Collins effect} [4,5].  Here, a
transfer of polarization has to occur
not only from the proton to quark $q$ but also from $q$ to $q'$.

\bigskip
\noindent
{\sl 3.2 The Collins effect.}

\smallskip
\noindent
According to Collins [4,5], the fragmentation function of the transversely
polarized quark $q$ takes the form
$$ D_{\pi/q}(\vPol_q,z, \hT) =
\bar D_{\pi/q}(z, \hT)
\left\{ 1 + {\cal A}_{\pi/q} (z, \hT) \vert\vPol_q\vert
\sin[\varphi (\vPol_q)-\varphi (\vhT)] \right\}
\,,\eqno(3.2)$$
where $\vPol_q$ is the quark polarization vector ($|\vPol_q| \le 1$), $\vec
h_{\T}$ is the pion transverse momentum relative to the quark momentum $\vec q$
and $\varphi (\vec a)$ is the azimuth of $\vec a$ around $\vec q$.  The factors
after ${\cal A}$ can be replaced by $|\vec q \times \vhT |^{-1} \ \vPol_q \cdot
(\vec q \times \vhT)$.  Such a dependence is allowed by P- and T- invariance
but
still remains to be measured.

The Collins effect is the reciprocal of the Sivers effect.  However, Collins
argued that the latter is prohibited by time reversal invariance [4], while the
former is not.  As for the mechanism b), it vanishes for massless quarks due to
chiral symmetry:  single spin asymmetry in $q \rightarrow q^{\prime}$ is not
compatible with conservation of quark helicity.  Therefore it should be small
for hard or semi-hard scattering at high energy.  In conclusion, among the
sources of asymmetry a), b) and c) discussed above, we have a preference for
the
Collins effect illustrated by Fig.~1.\footnote{**}{
   We shall not discuss other approaches [14] not relying on the factorized
   parton description (Eq.~(3.1) or (3.3)).  They are not necessarily
   in contradiction with the present one.}

\bigskip
\noindent
{\sl 3.3 Consequence for the single spin asymmetry.}

\smallskip
\noindent
Let us consider the hypothesis that the E704 asymmetry is due to the
Collins effect. In the parton model, the polarized inclusive cross section
reads
$${d\sigma\over d^3\vec p} = \sum_{\rm a,b,c,d} \int dx
q_a(x) \int dy q_b (y)\ \times $$
$$\int d\cos\hat\theta\, d\hat\varphi \, {d\hat\sigma^{q_a+q_b\to q_c+q_d}\over
d\hat\Omega} \int dz\, d^2 \hT\ D_{\pi/q_c} (\vPol_{q_c},z,\vhT) \, \delta(\vec
p-z\vec c-\vhT),
\eqno(3.3)$$
where $\vec c$ is the momentum of the scattered quark.
The transverse polarization of that quark is given by
$$\vPol_{q_c} = {\cal R} \ \vPol_{\rm beam} \ {\Delta_{\T}q_a(x) \over q_a(x)}
\ \hat D_{NN} (\hat\theta) \,.
\eqno(3.4)$$
${\cal R}$ is the rotation about $\vec p_{\rm beam} \times \vec c$ which
brings $\vec p_{\rm beam}$ along $\vec c$,
$$\Delta_{\T} q(x) \equiv q\!\uparrow(x) - q\!\downarrow(x)\, ,
\eqno(3.5)$$
also called $h_1(x)$,
is the quark transversity distribution [15,16] in the proton polarized
upwards, and $\hat D_{NN}(\hat\theta)$ is the coefficient of the spin transfer,
normal to the scattering plane, in the subprocess.  Formula (3.1) results from
integration of (3.3) over the momentum fraction $y$ of the parton of the
target.
$q_b$ and $q_d$ in (3.3) may also be replaced by gluons. This does not change
the results concerning the asymmetry.

At large $x_F$, the dominant quark flavours are $q_a=q_c=u$ for $\pi^+$
production and $q_a=q_c=d$ for $\pi^-$ production.  Furthermore, the hard
scattering occurs predominantly at small $\hat\theta$ and $\hat D_{NN}
(\hat\theta)$ is close to unity, as in the case of the $\hat t$-channel
one-gluon exchange where $\hat D_{NN}= -2 \hat s \hat u / (\hat s^2 + \hat
u^2)$; $\hat s$ and $\hat u$ being the Mandelstam variables of the parton
subprocess.  Thus, the results of E704 collaboration imply

$${\Delta_{\T} u(\bar x) \over u(\bar x) }
{\cal A}(\bar z,\bar h_{\T}) \ge {\rm about}\ 0.4
\,,~\eqno(3.6)$$

$${\Delta_{\T} d(\bar x) \over d(\bar x) }
{\cal A}(\bar z,\bar h_{\T}) \le {\rm about}\, -0.4
\,,\eqno(3.7)$$
for $\bar x \bar z \simeq x_F \simeq 0.8$.  $\bar x$ means the most probable
value of $x$.  The inequalities take into account the fact that integration
over
$\vqT$ and $\hat\theta$ always dilutes the Collins asymmetry.  It means that we
get at least a lower bound of 0.4 separately for $|\Delta_{\T} u / u|$ ,
$|\Delta_{\T} d / d|$ at large $x$ and $|{\cal A}(z,h_\T)|$ at large $z$ and
for
the most probable value of $h_\T$.

\bigskip
\medskip
%**************************************************************************
\noindent
{\bf 4. Simple model of single-spin asymmetry}

\smallskip
\noindent

In order to make the conclusions of the previous section more quantitative, we
have constructed a simple model based on the string model [17] of quark
fragmentation.

We consider only the valence quarks of the projectile proton.  We assume that
the quark elastic-scattering cross-section ${d\hat\sigma / dq_\T}$ in (3.3)
depends only on the transverse momentum $q_\T$ of the scattered quark.  Since
the scattering angle is in our case very small we assume $D_{NN} =1$ in
Eq.~(3.4).  After scattering, the quark spans a string between itself and the
target.  The string decays according to the recursive Lund recipe [17], for
which we use the Standard Lund splitting function:
$$
f(z)=(1+C)(1-z)^C \,,
\eqno(4.1)$$
$z$ being the fraction of the null plane momentum $p^+ \equiv p^0+p^3$ of the
string taken away by the next hadron.  $f(z) = D^{\rm rank=1}(z)$ corresponds
to the production rate of the leading\footnote{***}{We call "leading" or
"first-rank" the hadron which contains the original quark spanning the string.}
hadron.  This gives for all ranks the fragmentation function [17]:

$$
D(z) = (1+C) {1 \over z} (1-z)^C = {1 \over z} f(z).
\eqno(4.2)$$
Thus, for all the other (subleading) hadrons originating from
the string we get:

$$
D^{{\rm rank}\, \ge\, 2}(z) = (1+C) {1-z \over z} (1-z)^C = f(z){1-z \over z}.
\eqno(4.3)$$
The transverse momenta, relative to the direction of the string, of the quark
($\vec k_{q\T}$) and the antiquark ($\vec k_{\bar q\T}$) of every pair created
in the string balance each other (local compensation of the transverse
momentum)
and are distributed according to

$$
\rho(\vec \kT)\, d^2\vkT =
{d^2\vkT \over \kappa} \exp \left({-\pi \kT^2 \over \kappa}\right)\,,
\eqno(4.4)$$
$\kappa$ being the string tension.

Each quark-antiquark pair created during the string breaking is assumed to be
in
a $^3P_0$ state (vacuum quantum numbers) [18], {\it i.e.,} with parallel
polarizations.  According to the Lund mechanism for inclusive $\Lambda$
polarization [17], their polarizations are correlated to the transverse
momentum
$k_{\bar q\T}$ of the antiquark by

$$
\vPol_{\bar q}(\vec k_{\bar q\T}) = \,-\,{L \over \beta+L}
{\hat z \times \vec k_{\bar q\T}
\over  k_{\bar q\T}} \,,
\eqno(4.5)$$
where $\hat z$ is the unit vector along the $z$ direction, $\beta$ is the
parameter determining the correlation and $L$ is the classical orbital angular
momentum of the $q\bar q$ pair:
$$
L = { 2\ k_{\bar q\T} \sqrt{m_q^2 + k_{\bar q\T}^2} \over \kappa }
  \simeq { 2\ k_{\bar q\T}^2 \over \kappa }
\eqno(4.6)$$
(see Fig.~2 for the schematic explanation).

In order that $q_0$ and $\bar q_1$ of Fig.~2 combine into a pion, they have to
form a spin singlet state, the probability of which is
$$
{1 \over 4} \, (1 - \vPol_{q_0} \cdot \vPol_{\bar q_1}) \,,
\eqno(4.7)$$
in accordance with the projection operator on the singlet state
${1 \over 4} - \vec s(q_0) \cdot  \vec s(\bar q_1)$, $\vec s$ being the quark
spin operator.  The polarization of the leading quark $q_0$ is
$$
\vPol_{q_0}(x) = {\Delta_{\T}q_0(x) \over q_0(x)} \cdot \hat y.
\eqno(4.8)$$
The factor (4.7) causes the Collins effect:  if $q_0$ in Fig.~2 is polarized
upwards then $\bar q_1$  --- and the pion which contains $\bar q_1$ --- tends
to go to the left-hand-side of the $\hat z$ direction.

Putting together Eqs~(4.1), (4.4) and (4.7), we get the contribution of
the leading pions to the fragmentation function of the polarized quark $q$:
$$
D^{\rm rank=1}_{\pi/q}(z, \vhT) =
c_1 D^{\rm rank=1}(z) \rho (\hT) (1 - \vPol_q \cdot \vPol_{\bar q}(\vhT)),
\eqno(4.9)
$$
where $\vPol_{\bar q}$ is given by (4.5) and (4.6).
$c_1$ is the probability that the flavours of $q$ and $\bar q$ combine into the
pion of the appropriate charge.  We do not take into account the vector mesons
since at high $x_F$ the pions are mostly produced directly.  Therefore we omit
the factor ${1\over4}$ of Eq.~(4.7).

We do not introduce the Collins effect in subleading ranks.
The subleading fragmentation function is spin independent and reads:
$$
D^{{\rm rank}\, \ge\, 2}_{\pi/q}(z, \vhT) =
c_2 D^{{\rm rank}\, \ge\, 2}(z) \rho_\pi (\hT),
\eqno(4.10)
$$
where $\rho_\pi (\hT)$ is the distribution of the transverse momentum of
the produced pion with respect to the direction of the fragmenting quark.
It is the convolution of the transverse momentum distributions
of its constituents (4.4) and is also a Gaussian function but of the twice
larger variance. $c_2$ is the flavour factor analogical to that of Eq.~(4.9).

We do not take into account the second string which is spanned by the
remnant diquark of the projectile. A large part of the energy of that
string goes into the leading baryon and it does not contribute much
to the pion spectrum at high $x_F$.
\vfill
\eject

Our final formula for the polarized cross-section reads:
$$
{d\sigma \over dx_F d^2\vpT} =
\sum_{q=u,d}\int dx q(x) \int d^2\vqT {d\hat\sigma \over d^2 \vqT}
\int dz d^2\vhT D_{\pi/q}(z,\vhT) \times
$$
$$
\delta\left(x_F - \sqrt{z^2x^2-{4\pT^2\over s}}\right)
\delta^2(\vpT - z \vqT - \vhT),
\eqno(4.11)
$$
where
$$
D_{\pi/q}(z,\vhT) =
D^{\rm rank=1}_{\pi/q}(z, \vhT) + D^{{\rm rank}\, \ge\, 2}_{\pi/q}(z, \vhT).
\eqno(4.12)
$$

Since the production rate varies with the azimuthal angle $\phi$ of the pion
momentum like in (3.2) then, in order to obtain the asymmetry at given values
of
$\pT$ and $x_F$, we need to compare $d\sigma (x_F,\pT,\phi)$ only at $\phi = 0$
and $\phi = \pi$.  Thus, for the transverse spin asymmetry we get:
$$
A_N(x_F,\pT) = {d\sigma (x_F,\pT,0) - d\sigma(x_F,\pT,\pi) \over
                d\sigma(x_F,\pT,0) + d\sigma(x_F,\pT,\pi)}
\eqno(4.13)$$.

\bigskip
\medskip
%**************************************************************************
\noindent
{\bf 5. Numerical results}

\smallskip
\noindent
For the numerical calculations we parametrized the quark distributions as
follows:
$$
u(x) = {16\over 3\pi}\ x^{-1/2}\ (1-x)^{3/2}
$$ $$
d(x) = {15\over 16}\ x^{-1/2}\ (1-x)^2.
\eqno(5.1)
$$
One has to note that the scale of the process we are dealing with here is
usually below 1GeV. This means that one cannot use the quark
distributions obtained from the deeply inelastic scattering at high
$Q^2$. In this region we have to use a parametrization being between the
large $Q^2$ region, where, at high $x$, $u(x)\sim(1-x)^3$ and $d(x)\sim(1-x)^4$
and the dual parton model region, where the quark-diquark splitting
function, $q(x)\sim(1-x)$.

We have used the string tension $\kappa = 0.197$ GeV$^2$ (it corresponds to
1GeV/fm) and the parameter of the
fragmentation function $C = 0.5$. In pair creation we have used the flavour
abundances with the ratio $u :  d :  s = 3 :  3 :  1$, which determines the
coefficients in Eqs~(4.9) and (4.10)  to be $c_1 = 3/7$, $c_2 = 9/49$ for
charged and $c_2 = 18/49$ for neutral
pions\footnote{$^{\S}$}{
In this model, every $u\bar u$ or $d\bar d$ meson is considered as a $\pi^0$
(no
$\eta^0$) ; it gives $\sigma(\pi^+)+\sigma(\pi^-) = \sigma(\pi^0)$, instead of
$2\ \sigma(\pi^0)$ as required by isospin.  Nevertheless, Eq.~(5.8) below
remains true.}.

The quark scattering cross-section was parametrized as
$$
{d\sigma\over d^2\qT} \sim {1\over (\qT^2 + M^2)^3}
\eqno(5.2)
$$
with the parameter $M^2 = 0.5$GeV$^2$ making this cross-section normalizable.

Such a parametrization, with the power-law decrease advocated by Field and
Feynman [19], gives, up to the overall normalization, a good agreement
with the experimental inclusive spectra of pions. We show the comparison to the
400GeV and 360GeV experimental data [20,21] in Fig.~3.

As it was noted in section 2, if one assumes the Collins effect to be
responsible for the observed asymmetry, then the E704 data implies
$\Delta_\T u / u = -\Delta_\T d / d$ at high $x$ thus indicating a violation
of $SU(6)$, where $\Delta_\T u / u = 2/3$ and $\Delta_\T d / d = -1/3$.
For the numerical calculations of the single spin asymmetry, we assumed that
$$
\Delta_\T u / u = -\Delta_\T d/d = \Pol(x) = \Pol_{\max} x^n.
\eqno(5.3)
$$
We found that $\Pol_{\max} = 1$ and $n=2$
gives the best agreement with the experimental $x_F$ dependence of the
asymmetry.

Such a transversity does not violate the positivity constraints
derived recently by Soffer [22]. The Soffer's inequality relates the
transversity distribution $\Delta_\T q = h_1$ to the helicity distribution
$\Delta q = g_1$ and reads:
$$
2|\Delta_\T q| \le q + \Delta q.
\eqno(5.4)
$$
If one takes into account only the valence quarks, then the helicity
asymmetries of the proton, measured in the deeply inelastic scattering, is:
$$
A_1^p = {4 \Delta u + \Delta d \over 4 u + d}
\eqno(5.5)
$$
and that of the neutron equals:
$$
A_1^n = {\Delta u + 4 \Delta d \over u + 4 d}.
\eqno(5.6)
$$
Thus, if we assume that $\Delta_\T u / u = -\Delta_\T d/d = \Pol(x)$, then
the Soffer's inequality implies
$$
A_1^p(x) \ge 2 \Pol(x) - 1,
$$ $$
A_1^n(x) \ge 2 \Pol(x) - 1.
\eqno(5.7)
$$
The above inequalities are satisfied by the present data if one assumes
$\Pol(x) = x^2$. This is shown in Fig.~4 where $A_1^p(x)$ measured
by SMC [23] and E143 [24] and $A_1^n(x)$ measured by the E142 collaboration
[25] are plotted together with the curves representing Eqs~(5.7).
One can see that $\Pol(x) = x^2$ is well within the limits and $\Pol(x) = x$
is still allowed by the data, as well as the powers of $x$ higher than 2.

We plot the results of our calculation of the transverse spin asymmetry in
Figs.~5--8. The full lines and the dashed ones in Figs~7 and 8 show the
predictions of our model obtained with the parameter $\beta$ of Eq.~(4.5)
equal 1 [17].

These predictions are in reasonable agreement with most of the data.
Only the asymmetry measured by the E704 collaboration at 0.7GeV$<\pT <$2GeV
(Fig.~5a) is strongly underestimated. Our model gives the opposite asymmetries
for $\pi^+$ and $\pi^-$ and predicts the increase of the absolute values
of the asymmetries with $x_F$. Nevertheless, it cannot accout for the very
strong $\pT$ dependence of the asymmetries measured by E704. We got
agreement with the low-$\pT$ data but underestimate the high-$\pT$ ones.

However, the $\pT$ dependence of the E704 data was discussed recently by
Arestov [26] from the purely experimental point of view and was found to
be questionable. Also the asymmetries of $\pi^0$, measured
by E704 in the central region, showed initially a very strong $\pT$ dependence
[10] but after reanalysis [11] showed no such dependence
at all and are consistent with zero. Both the lack of the strong
$\pT$ dependence of the asymmetry and its very small magnitude in the central
region are in agreement with the Collins effect.

In the previous version of this paper [27] we were able to obtain
the strong $\pT$ dependence of the asymmetry at high $x_F$ but it
was only due to the fact that we neglected the quark scattering and
assumed only the exponentially falling intrinsic $\qT$ distribution
of the leading quark, similar to the $\hT$ distribution in the
fragmentation. In the present approach, the
$\qT$ distribution (5.2) in quark scattering is much flatter than the
$\hT$ distribution (4.4). Thus, at high $\pT$ the
contribution of the transverse momentum of the fragmentation (which
determines the asymmetry) to the
total $\pT$ saturates. This
makes the asymmetries rise at rather low $\pT$ and then flatten at
higher $\pT$.

We checked, by forcing $\beta$ in (4.5) to be 0, that the strong enough
Collins effect (satisfying the inequalities (3.6) and (3.7)) can account for
the magnitude of the experimental asymmetries
of Fig.~5a. Nevertheless, $\beta = 0$ (100\% spin-$\kT$ correlation
in string breaking independent of $\kT$) is not thinkable. It does
not change the $\pT$ dependence either; so large Collins effect leads to
strong overestimation of the lower-$\pT$ data.

As regards the region where both $x_F$ and $\pT$ are low (small $x_F$
points of Fig.~5b), the calculated asymmetry can be slightly
overestimated. At so low $p_\T$ and rather small $x_F$, the pions produced
in the decay of the second string, spanned by the diquark, can contribute
significantly and wash out the asymmetry.

The asymmetry of $\pi^0$ is a combination of the $\pi ^+$ and $\pi^-$ ones:
$$ A_N(\pi^0) = {\sigma(\pi^+)\ A_N(\pi^+) + \sigma(\pi^-)\ A_N(\pi^-)
\over \sigma(\pi^+) + \sigma(\pi^-) } \,.
\eqno(5.8)
$$
This relation follows from the isospin symmetry for
isoscalar targets. For the proton target it holds provided that the
isospin correlations are of short range in rapidity; this is the
case in the multiparticle production.
Nevertheless, we show in Fig.~6 the comparison of our results to the E704 data
[7,8] on $\pi^0$ production in $pp$ and $\bar pp$ collisions.
Here there is no discrepancy between the model and the data.
Note that our model predicts the same
asymmetry for the proton and the antiproton beams.

For completeness we show in Fig.~7 the earlier data measured with 13.3 and 18.5
GeV polarized protons [9]. Here the agreement of the model and the data
is also good. Only the asymmetry of $\pi^-$ calculated in the model
tend to overestimate the data, particularly at low $x_F$.
However, the $\pi^-$ data has been measured at very low $p_\T$ and the
contribution of the diquark fragmentation can be not negligible also here.

Finally, in Fig.~8 we show the predictions of the model for the asymmetry
of charged kaons. Since in this model the asymmetry is a purely leading
effect, the asymmetry of $K^-$ vanishes. Measuring this asymmetry and its
$x_F$ dependence would provide information on whether and to what extent
the asymmetry is limited to the leading particle. From the point of view
of our model, the nonvanishing asymmetry of $K^-$ would be an indication
of the Collins effect in higher-rank hadrons.

The asymmetry of $K^+$ is predicted to be similar to that of the positive
pions.

\bigskip
\medskip
%**************************************************************************
\noindent
{\bf 6. Discussion and conclusions}

\smallskip
\noindent
To summarize, we have calculated the single transverse spin asymmetry in
high-energy $pp$ collisions in a simple model involving the Collins effect
(asymmetry arising at the level of fragmentation of a quark into hadrons).
We parametrized the Collins effect by the Lund mechanism of polarization in
the coloured string model.

We got qualitative agreement with the data when we assumed that:

\item{a)}
The transverse polarization of the $u$ and $d$ quarks in the transversely
polarized proton are close to unity but of the opposite sign ($\vPol_u =
\vPol_{\rm proton}$, $\vPol_d = - \vPol_{\rm proton}$) at momentum fraction $x$
close to 1.

\item{b)}
The dependence of the quark transversity (or polarization) on the momentum
fraction $x$ is close to be proportional to $x^2$.

\noindent
However, the Collins effect cannot explain the very strong $\pT$ dependence of
the E704 data. Some additional mechanism of the asymmetry would be needed
in order to account for the E704 data for charged pions at $\pT > 0.7$GeV.
Apart of this set, our model gives good agreement with the data.
Presently, we do not find any mechanism which could remove the above
discrepancy.

The quark transversities we inferred at large $x$:
$$
{\Delta_{\T} u(x)\over u(x)}\approx
\,-\,{\Delta_{\T} d(x)\over d(x)} \to 1 \qquad ({\rm for}\ x\to 1)
\eqno(6.1)$$
are, in fact, not unreasonably large. Consider a covariant model of the
baryon consisting of a quark and a bound spectator diquark [16,28] ;
then
$$
q\uparrow(x)= {x \over 16\pi^{2}}
\int_{-\infty}^{q^2_m} dq^2  \left({g(q^2) \over
q^2-m^2_q}\right)^2
\sum_{\rm diquark\ polarization}
\left|\bar u(q \uparrow) V u(p \uparrow)\right|^2
\eqno(6.2)$$
where $g(q^2)$ is the $q-q\!q-B$ form factor, $V=1$ for a scalar diquark,
$V=\gamma_5 \gamma \cdot \varepsilon$\  for a $1^+$ diquark of polarization
$\varepsilon ^\mu$ and
$$
q^2_m= xm^2_B-{x \over 1-x} m^2_{q\!q} \,.
\eqno(6.3)$$
Formula (6.2) is similar to the covariant Weizs\"acker--Williams formula, but
for a ``spin ${1 \over 2}$ cloud'').  Independently of $g(q^2)$, this model
predicts the following behaviour at $x\to 1$:

\item{-}
for a $1^+$ spectator diquark, helicity is fully transmitted (~$\Delta_L q(x)
/q(x)\to 1$~), transversity is fully reversed (~$\Delta_{\T} q(x) / q(x)
\to -1$~).  In particular, $\Pol_d(x) \to -1$.

\item{-}
for a $0^+$ diquark, $\Delta_{\T} q(x)$ and $q^+(x)$ coincide and, for
$g(q^2)$ decreasing faster than $q^{-2}$, they exceed ${2\over3} q(x)$ as
$x\to 1$

\noindent
Thus, a dominance of the scalar spectator for the $u$ quark and the
pseudo-vector one for the $d$ at $x\sim 1$ could lead to the large opposite
transversities as in (6.1).

The conclusion b), related to the $x_F$ dependence, is model-dependent and
does not need to be considered as a firm prediction.  One needs a good
parametrization of the Collins
effect before one can deduce the $x$ dependence of the quark transversity.  Our
parametrization is an approximation which should work only at reasonably high
values of $x_F$.  We took into account the Collins effect only
for the first-rank (leading) hadrons, wherefrom ${\cal A} \propto z$ in (3.2).
In the string model,
the second-rank hadrons have the asymmetry of the opposite sign as compared to
the first-rank ones.  More generally, the subsequent ranks are asymmetric in
the
opposite way to each other (as required also by local compensation of
transverse
momentum).  This should cause a faster decrease of ${\cal A}$ at lower $z$
values, where the higher-rank hadrons are more important.  Unfortunatly this
feature was not possible to include in our simple semi-analytical calculation,
since the yields of rank-2 (and higher) hadrons do not have simple analytical
forms.  Assuming $\bar x\sim \bar z\sim \sqrt{x_F}$, a steeper $ {\cal A}$ (for
instance $\propto z^2$) would have to be compensated by a flatter $\Pol_q(x)$
(for instance $\propto x$).

The contribution of vector mesons also should reduce ${\cal A}$ at lower
$z$.  Moreover, it has been shown that the vector meson can also have
a {\it tensor} polarization [29] which would result in the Collins effect for
the decay products. We did not include this possibility. Another mechanism of
asymmetry can be the interference between direct and resonance production [30].

The main conclusion of this paper is that the single spin asymmetry may be the
first experimental indication for the existence of the Collins effect.  A more
detailed experiment would be useful to select between this and alternative
explanations.  Besides its theoretical interest, the Collins effect may be the
most efficient "quark polarimeter" necessary for the measurements of the
transversity distributions in the nucleons [5,31].  We hope that this effect
will soon be tested directly, for instance in the azimuthal correlation of two
pion pairs from opposite quark jets in $e^+ e^-$ annihilation.

\bigskip
\medskip
\noindent
%**************************************************************************
{\bf Acknowledgements}
\smallskip

\noindent
We are grateful to J.~Szwed for discussions. X.A. and J.C. acknowledge the
financial support from the IN2P3--Poland scientific exchange programme.
J.C. has been also supported by the Polish Government grants of KBN no.\
2~0054~91~01, 2~0092~91~01 and 2~2376~91~02 during completion of this
work.

\bigskip
\medskip
\noindent
%**************************************************************************
\vfill
\eject
{\bf References}
\smallskip

\item{[1]} K.~Heller, $7^{\rm th}$ Int.\ Conf.\ on Polarization Phenomena in
Nuclear Physics, Paris 1990, p.\ 163, and references therein; P.R.~Cameron {\it
et al.}, \PR {\bf D32}, 3070 (1985); T.A.~Armstrong {\it et al.}, \NP
{\bf B262}, 356 (1985); S.~Gourlay {\it et al.}, \PRL {\bf 56}, 2244 (1986);
M.~Guanziroli {\it et al.}, \ZP {\bf C37}, 545 (1988)

\item{[2]} D.~Sivers, \PR {\bf D41}, 83 (1990); \PR {\bf D43}, 261 (1991)

\item{[3]} J.~Szwed, Proc.\ of the 9$^{\rm th}$ International Symposium ``High
Energy Spin Physics'' held at Bonn, 6--15 Sep.\ 1990, Springer Verlag 1991;
\PL {\bf B105}, 403 (1981)

\item{[4]} J.~Collins, \NP {\bf B396}, 161 (1993)

\item{[5]} J.~Collins, S.F.~Heppelmann and G.A.~Ladinsky, \NP {\bf B420}, 565
(1994)

%----------- pi+ and pi-, fragmentation region ---
%
\item{[6]} E704 Coll., D.L.~Adams {\it et al.}, \PL {\bf B264}, 462 (1991)

%------------ pi0, fragmentation region -----
%
\item{[7]} E704 Coll., D.L.~Adams {\it et al.}, \ZP {\bf C56}, 181 (1992);
E704 Coll., B.E.~Bonner {\it et al.}, \PRL {\bf61}, 1918 (1988)

%------------ pbar->pi0 and p->pi0, fragmentation region -----
%
\item{[8]} E704 Coll., D.L.~Adams {\it et al.}, \PL {\bf B261}, 201 (1991)

%----------fragmentation region 13.3, 18.5 GeV
%
\item{[9]}
B.E.~Bonner {\it et al.}, \PR {D41}, 13 (1990);

%------------------pi0, central region -----
%
\item{[10]} E704 Coll., D.L.~Adams {\it et al.}, \PL {\bf B276}, 531 (1992)

%------------------pi0, central region, reanalized ----
%
\item{[11]} E704 Coll., D.L.~Adams {\it et al.}, Protvino preprint IHEP-{\bf
94-88}, IFVE-{\bf 94-88} (1994)

%----------central region 13.3, 18.5, 24., 40.GeV -----
%
\item{[12]} V.D.~Apokin {\it et al.}, \PL {\bf B243}, 461 (1990);
M.S.~Amaglobeli {\it et al.}, Sov.\ J.\ Nucl.\ Phys.\ {\bf 50}, 432 (1989);
J.~Antille {\it et al.}, \PL {\bf B94}, 523 (1980);
S.~Saroff {\it et al.}, \PRL {\bf 64}, 995 (1990)

\item{[13]} M.~Anselmino, M.~Boglione, F.~Murgia, INFN Cagliari preprints
DFTT-{\bf 47-94} and DFTT-{\bf 48-94} (1994)

\item{[14]} M.G.~Ryskin, Sov.\ J.\ Nucl.\ Phys.\ {\bf48}, 708 (1988); C.~Boros,
Liang Zuo-tang and Meng Ta-chung, \PRL {\bf70}, 1751 (1993); H.~Fritzsch, Mod.\
Phys.\ Lett.\ {\bf A5}, 625 (1990); P.G.~Ratcliffe, Proc.\ 10th International
Symposium on High Energy Spin Physics (Yamada Conference XXXV), November 9-14,
1992, Nagoya, Japan (edited by T.  Hasegawa, N.  Norikawa, A.  Masaike, S.
Sawada), Universal Academy Press, Inc.\ p.~635

\item{[15]} J.P.~Ralston and D.E.~Soper, \NP {\bf B152}, 109 (1979);
J.L.~Cortes, B.~Pire and J.P.~Ralston, \ZP {\bf C55}, 409 (1992);
R.L.~Jaffe and Xiangdong Ji, \NP {\bf B375}, 527 (1992)

\item{[16]} X.~Artru and M.~Mekhfi, \ZP {\bf C45}, 669 (1990)

\item{[17]} B.~Andersson, G.~Gustafson, G.~Ingelman, T.~Sj\"ostrand, \PRep\
{\bf
97} 31 (1983)

\item{[18]} A.~Le~Yaouanc, L.~Oliver, O.~P\`ene and J.-C.~Raynal, {\it Hadron
Transitions in the Quark Model} (Gordon and Breach, 1988)

\item{[19]} R.D.~Field, R.P.~Feynman, \PR {\bf D15}, 2590 (1977)

\item{[20]} NA27 Coll., M.~Aguilar-Benitez {\it et al.}, \ZP {\bf C50}, 405
(1991)

\item{[21]} NA23 Coll., J.L.~Bailly {\it et al.}, \ZP {\bf C35}, 309 (1987)

\item{[22]} J.~Soffer, \PRL {\bf 74}, 1292 (1995)

\item{[23]} SMC Coll., D.~Adams {\it et al.}, \PL {\bf B329}, 399 (1994)

\item{[24]} E143 Coll., K.~Abe {\it et al.}, \PRL {\bf 74}, 346 (1995)

\item{[25]} E142 Coll., D.L.~Anthony {\it et al.}, \PRL {\bf 71}, 959 (1993)

\item{[26]} Yu.I.~Arestov, \PL {\bf B333}, 255 (1994)

\item{[27]} X.~Artru, J.~Czy\.zewski, H.~Yabuki, preprint {\bf TPJU-12/94}
(Cracow, 1994) and {\bf LYCEN/9423} (Lyon), hep-ph/9405426, unpublished

\item{[28]} H.~Meyer and P.J.~Mulders, \NP {\bf A528}, 589 (1991)

\item{[29]} Xiangdong Ji, \PR {\bf D49}, 114 (1994)

\item{[30]} J.C.~Collins and G.A.~Ladinsky, PSU/TH/{\bf 114} (hep-ph/9411444)

\item{[31]} X.~Artru, QCD and High Energy Hadronic Interactions
(Ed.~J.~Tr\^an Thanh V\^an, Editions Fronti\`eres, 1993), p.~47

\bigskip
\medskip
\noindent
%**************************************************************************
\vfill
\eject
{\bf Figure captions}

\smallskip

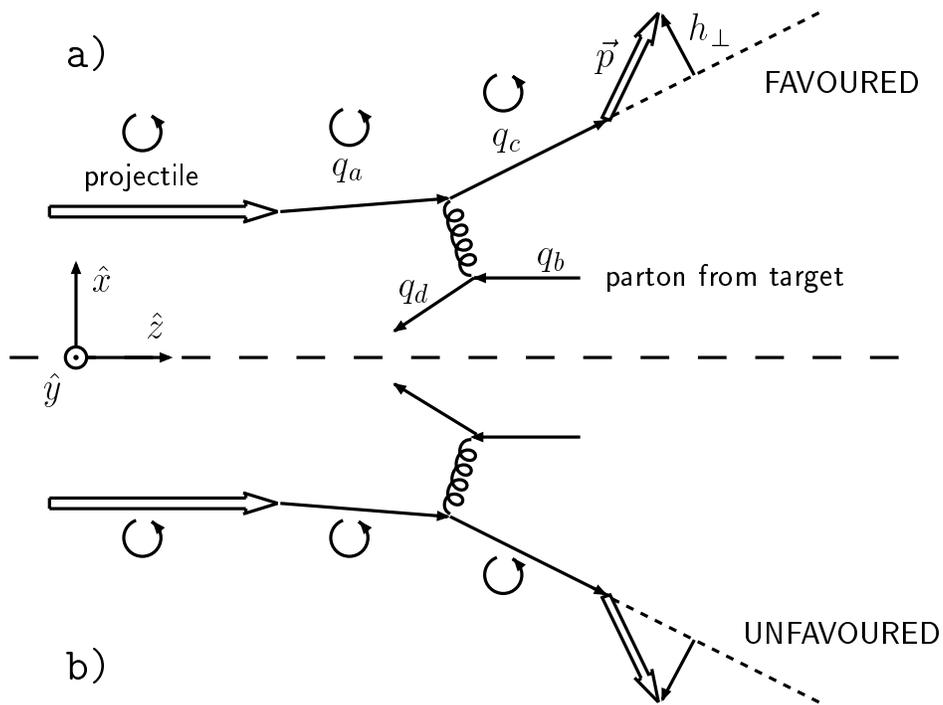
\item{Fig.~1} Inclusive pion production.  Two events (a) and (b), symmetric
with
respect to the $\hat y\hat z$ plane, are represented.  Without polarization,
they would have the same probability.  In the polarized case, the Collins
effect
favours the case (a).  The arrows labelled $q_i$ represent the momenta of the
quarks in the subprocess.  The spins are denoted by the arch-like arrows.  The
Collins effect acts at the last stage, where the quark $q_c$ fragments into the
pion carrying momentum $\vec p$.  $h_\T$ is the pion's transverse momentum with
respect to the quark $q_c$.

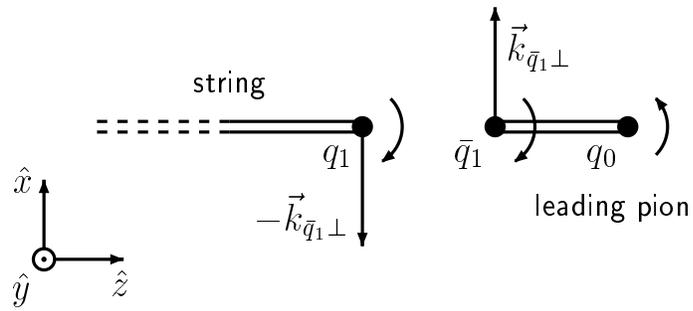
\item{Fig.~2} Production of the leading pion in a string spanned by
the transversely spinning quark $q_0$.

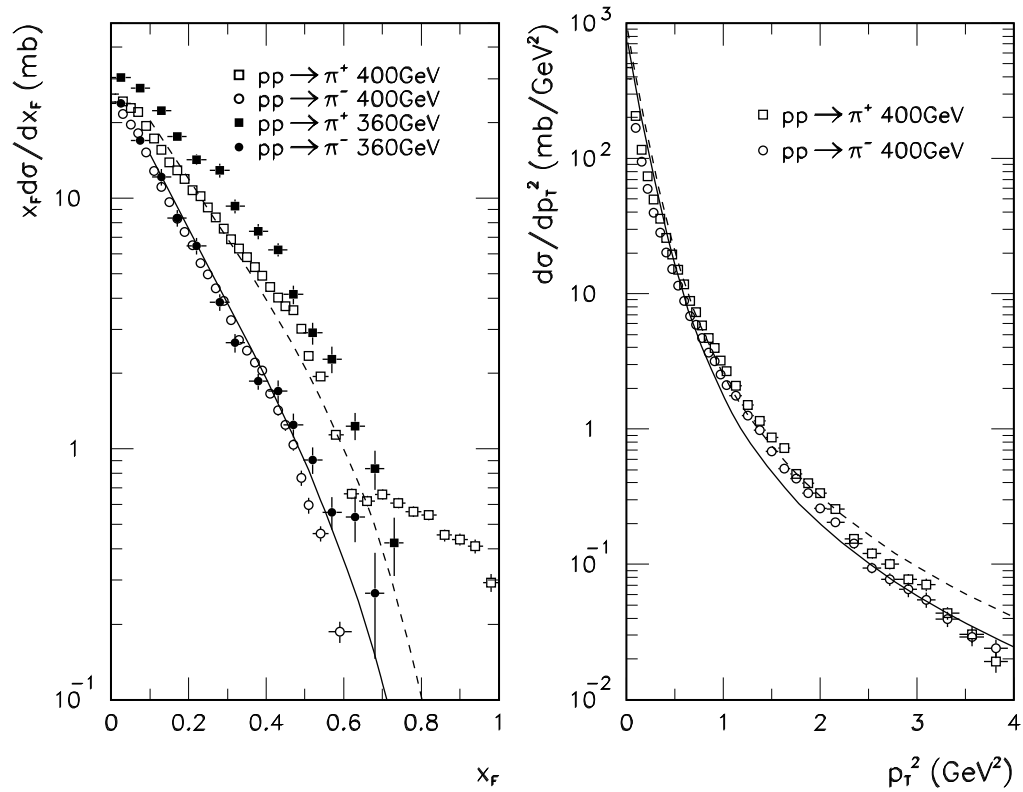
\item{Fig.~3} Inclusive cross-section for production of $\pi^+$ and $\pi^-$
plotted versus $x_F$ and $\pT^2$. The 360GeV data come from [21] and the
400GeV data from [20]. The dashed curve corresponding to $\pi^+$
and the solid one corresponding to $\pi^-$ show the results of
Eq.~(4.11) together with (5.1) and (5.2).

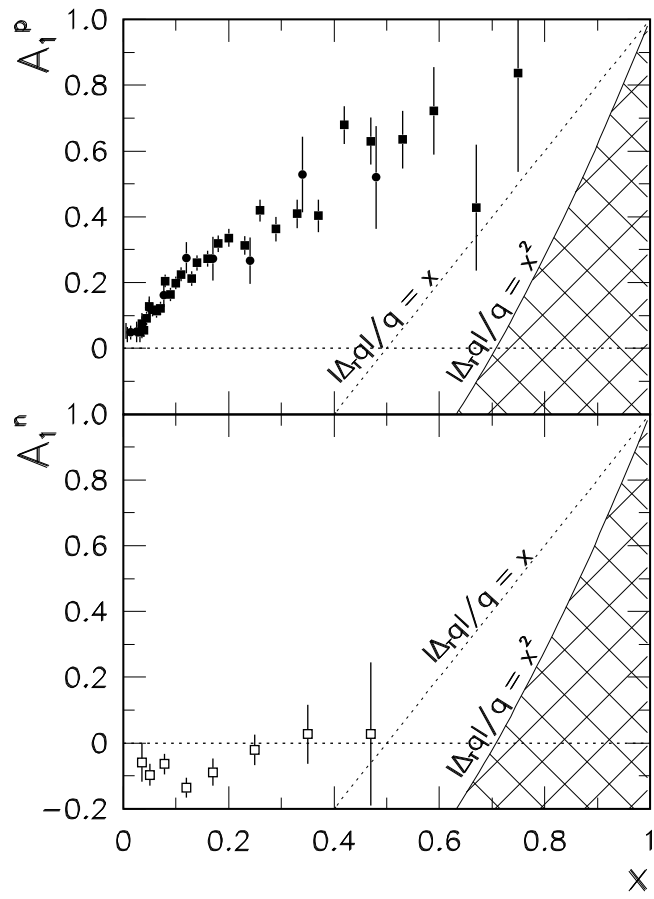
\item{Fig.~4} Longitudinal spin asymmetries of the proton and of the neutron
measured in deeply inelastic scattering.  The full squares are the E143 data
[24], the full circles the SMC data [23] and the open squares represent the
E142
data [25].  The values of $A_1^p$ and $A_1^n$ allowed by the Soffer's
inequality
(5.4) if $|\Delta_\T q|/q = x^2$ are denoted by the hatched area.  One can see
that $|\Delta_\T q|/q = x$ is also allowed by the present data.

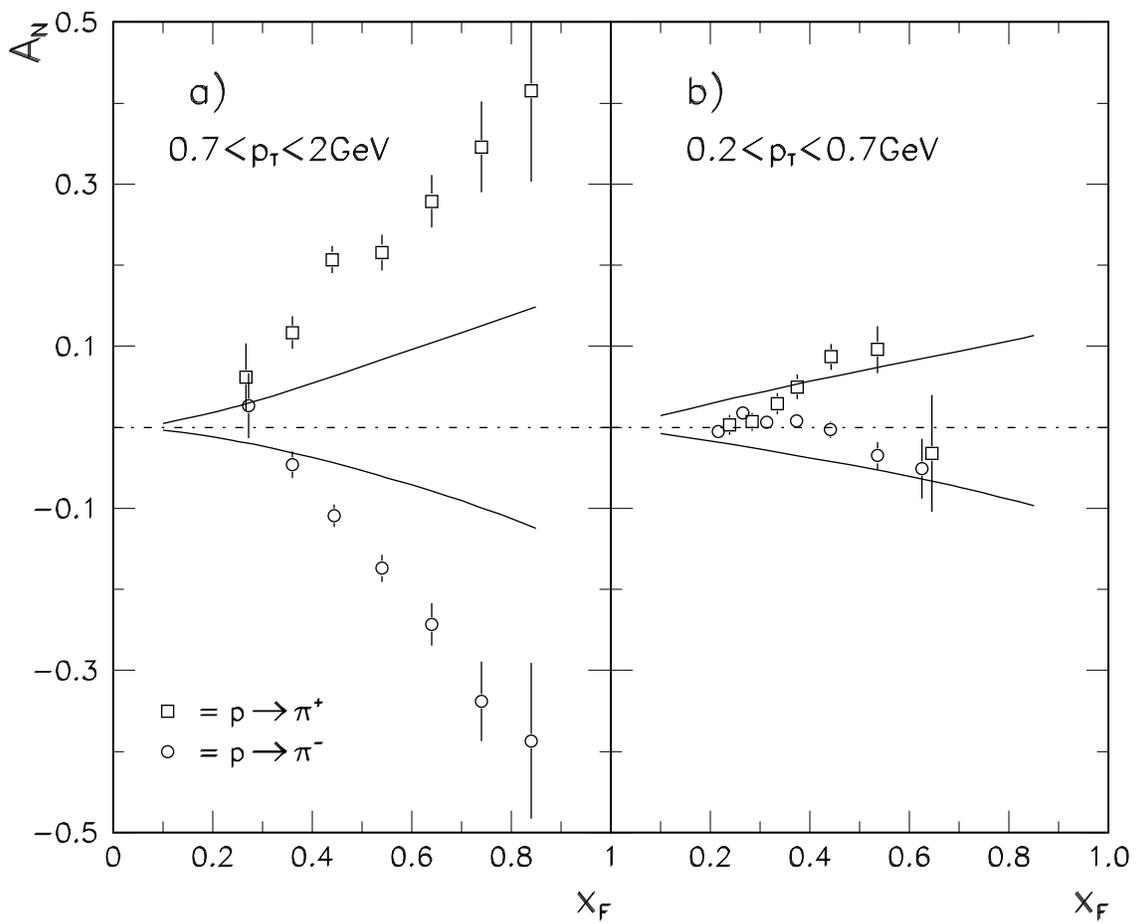
\item{Fig.~5} Single spin asymmetry measured by E704 collaboration for
charged pions at $0.2<p_\T<2.0$GeV [6].  The curves
are our model results calculated with quark transverse polarizations
$\Delta_\T u/u = -\Delta_\T d/d = x^2$ and $\beta = 1$.

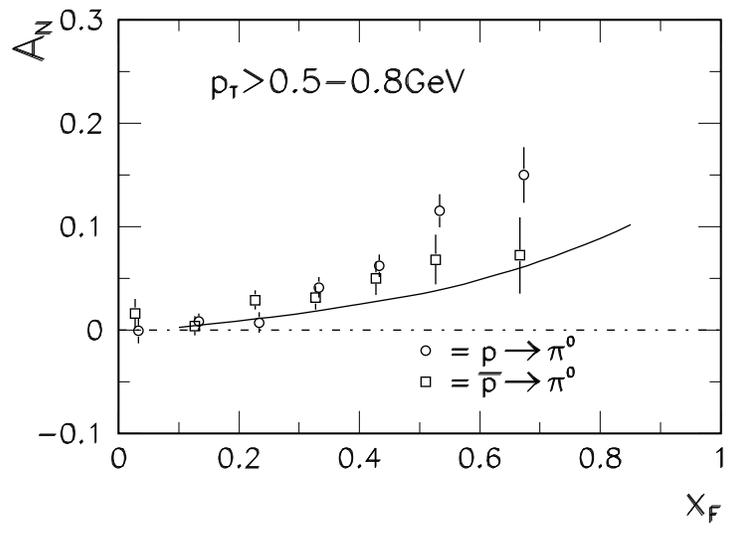
\item{Fig.~6} The asymmetry of $\pi^0$'s produced in $pp$ and $\bar pp$
collisions measured by the E704 collaboration [7,8].  The curve is our
model prediction.

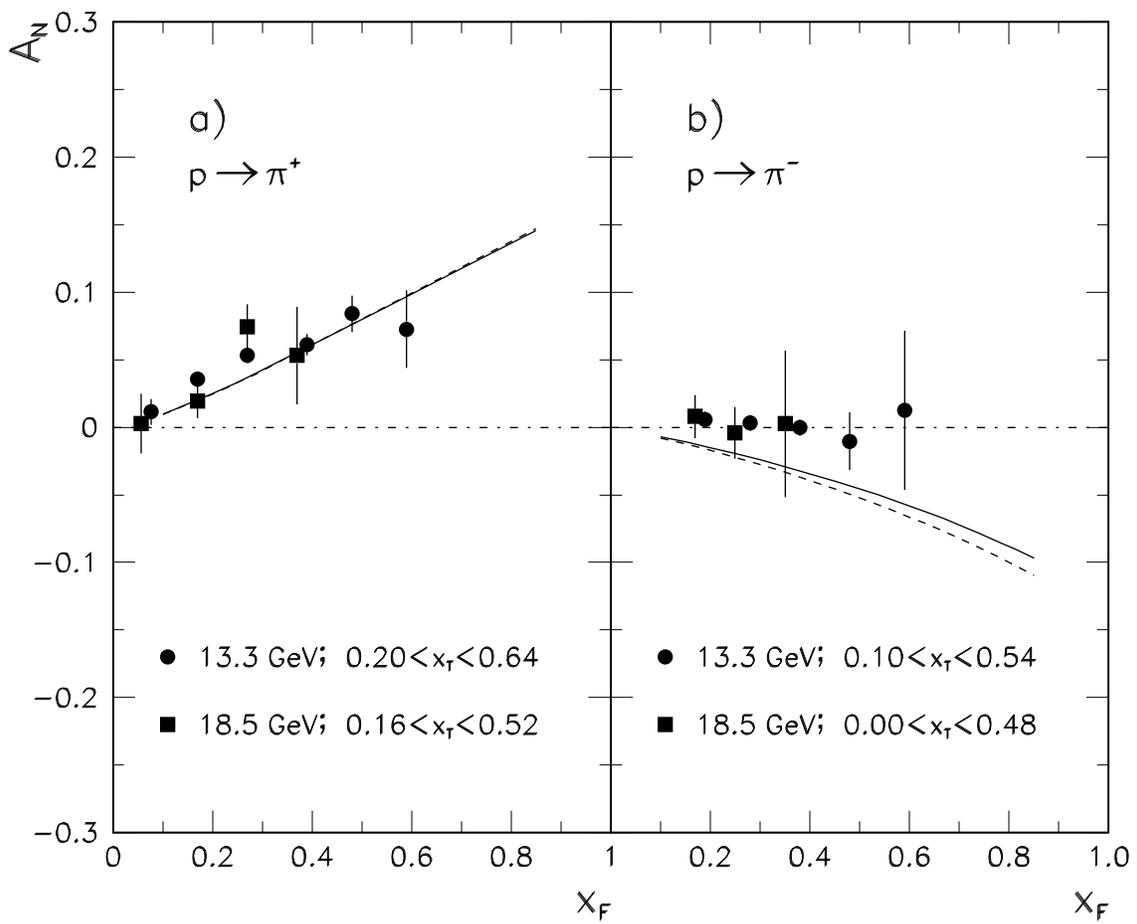
\item{Fig.~7} The single spin asymmetries of charged pions measured
with 13.3 and 18.5GeV proton beams [9]. The dashed lines are the
predictions of our model for 13.3GeV and the full ones are for 18.5GeV.

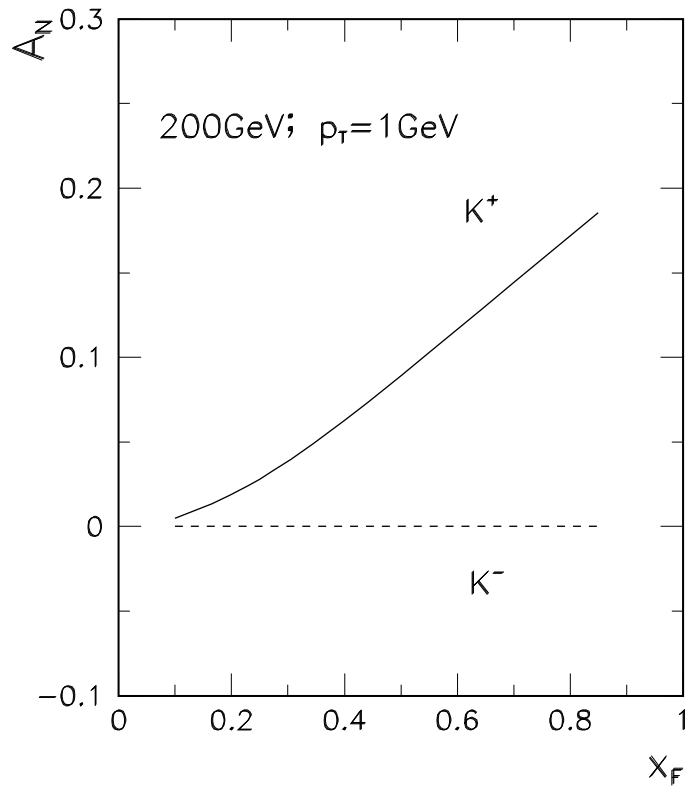
\item{Fig.~8} Transverse spin asymmetry of the charged kaons as predicted
by our model. No asymmetry is predicted for $K^-$ when the spin effects
are assumed to act only on the leading hadron.

\end